ARTICLE

# Developing an aqueous approach for synthesizing Au and M@Au (M = Pd, CuPt) hybrid nanostars with plasmonic properties†

Jingshan Du, Junjie Yu, Yalin Xiong, Zhuoqing Lin, Hui Zhang* and Deren Yang

Anisotropic Au nanoparticles show unique localized surface plasmon resonance (LSPR) properties, which make it attractive in optical, sensing, and biomedical applications. In this contribution, we report a general and facile strategy towards aqueous synthesis of Au and M@Au (M = Pd, CuPt) hybrid nanostars by reducing $HAuCl_4$ with ethanolamine in the presence of cetyltrimethylammonium bromide (CTAB). According to electron microscopic observation and spectral monitoring, we found that the layered epitaxial growth mode (*i.e.*, Frank-van der Merwe mechanism) contributes to the enlargement of the core, while, the random attachment of Au nanoclusters onto the cores accounts for the formation of the branches. Both of them are indispensable for the formation of the nanostars. The LSPR properties of the Au nanoparticles have been well investigated with morphology control *via* precursor amount and growth temperature. The Au nanostars showed improved surface-enhanced Raman spectroscopy (SERS) performance for rhodamine 6G due to their sharp edges and tips, which were therefore confirmed as good SERS substrate to detect trace amount of molecules.

## Introduction

Metallic nanoparticles (NPs) have received great attention due to their distinguished properties in catalysis,[1] optics,[2] and biomedicine.[3] Among those NPs of various metal elements and morphologies, anisotropic Au NPs have become extremely significant due to their unique localized surface plasmon resonance (LSPR) properties,[4,5] and thus widespread use in biological imaging, photothermal therapy and optical sensing.[6-8] The last decade has witnessed the significant progress in the synthesis of anisotropic Au NPs, with notable examples including nanorods,[9-12] nanostars,[13-15] and nanopods.[16,17] Such Au nanostructures show tunable LSPR peaks in the range from visible to near-infrared (IR) regions, which have been found mainly dependent on their aspect ratios. In addition, these anisotropic NPs can also create "hot spots" according to the coupling between their electromagnetic fields, which provide extremely large electric field enhancement around the nanostructures.[18,19] All of such fascinating characteristics facilitate the surface-enhanced Raman scattering where the Raman signals of the small molecules adsorbed on the NPs are significantly enhanced, thereby make it possible to sensing and detecting these molecules at trace amount level.[20]

Of these anisotropic nanostructures, Au nanostars with various branches are particularly interesting since their intrinsic properties result from hybridization of plasmon focalized at sharp edges and tips. Up to now, there are some synthetic routes that have been exploited for Au nanostars. To this end, Liz-Marzán *et al.* demonstrated the seed-mediated growth of Au nanostars in high yield through the reduction of $HAuCl_4$ in a concentrated solution of polyvinylpyrrolidone (PVP) in N,N-dimethylformamide (DMF).[14, 21-23] The use of PVP with a weak reducing power and DMF serving as a solvent played a key role in the formation of Au nanostars. In aqueous systems, a silver ion-induced strategy was also developed to generate Au nanostars.[13, 24-27] In these syntheses, the selective adsorption effect of silver ions on the surface of Au NPs through underpotential deposition (UPD) promoted their branched growth. Recently, a surfactant-assisted approach has been exploited for the synthesis of Au nanostars by taking advantages of simple procedure and high efficiency. For example, Pallavicini *et al.* reported the synthesis of branched Au NPs with the use of laurylsulfobetaine as the Zwitterionic surfactant.[28, 29] In another study, many researchers demonstrated successful synthesis of Au nanostars with various branched arms in the presence of complex amine molecules such as hydroxylamine,[30, 31] melamine,[32] hexamethylenetetramine (HMT),[16] and bis(amidoethyl-carbamoylethyl)octadecylamine ($C_{18}N_3$).[33] Although significant advances have been achieved on the synthesis of Au nanostars, the origin and mechanism for the formation of these branched structures have not been fully understood.

In parallel, bimetallic or multi-metallic NPs also show significant potential in catalysis or other multifunctional applications. Formation of core@shell bimetallic nanostructures has been considered as an efficient way to alter the electron structure of the shells due to a strong coupling between these two metals, which can significantly affect their catalytic performance and stability.[34-36] Au-based bimetallic NPs are also interesting candidates of plasmonic multifunctional colloids since they bring the unique optical response of Au to other functional metal NPs.[37] Therefore, developing a general and facile strategy for synthesizing anisotropic Au and Au-based bimetallic or multi-metallic nanostars, together with clarifying their formation mechanism, is still of great importance to the scientific community.

Herein, we report a general, facile, and powerful strategy towards aqueous synthesis of Au and M@Au (M = Pd, CuPt) hybrid nanostars in a controlled manner. Star-like Au shells were successfully grown on different as-preformed seeds including Au nanospheres, Pd nanocubes and CuPt bimetallic nanocubes in an aqueous solution containing $HAuCl_4$ and cetyltrimethylammonium bromide (CTAB) with ethanolamine as both a capping and reducing agent. Quantitative studies revealed that two concurrent mechanisms are responsible for the anisotropic morphology and the corresponding LSPR-induced absorption properties. The interior core of the NPs was gradually grown up due to the conventional layered epitaxial mechanism, *i.e.*, Frank-van der Merwe (F-M) growth mode. In contrast, the formation of the branches was attributed to the random attachment of Au nanoclusters onto the core particles. The as-prepared Au nanostars showed improved surface-enhanced Raman spectroscopy (SERS) performance compared to Au nanospheres in detecting trace amount of rhodamine 6G molecules.

## Experimental section

### Chemicals

Chloroauric acid ($HAuCl_4·4H_2O$, AR), cetyltrimethylammonium bromide (CTAB, AR), ethanolamine (AR), potassium chloride (KCl, AR) and hydrochloric acid (HCl, AR 36%-38%) were supplied by Sinopharm Chemical Reagent. L-ascorbic acid (AA, BioXtra, ≥99.0%), potassium tetrachloroplatinate(II) ($K_2PtCl_4$, 99.99% trace metals basis), sodium tetrachloropalladate(II) ($Na_2PdCl_4$, 99.99% trace metals basis), copper(II) chloride dihydrate ($CuCl_2·2H_2O$, reagent grade), and polyvinylpyrrolidone (PVP, MW=40,000) were supplied by Sigma-Aldrich. Sodium borohydride ($NaBH_4$, ≥98.0%) was supplied by Aldrich. Potassium bromide (KBr, AR) was supplied by Shanghai No. 4 Reagent. Ultrapure de-ionized water (DI water) was produced by a Milli-Q Synergy water purification system. All chemicals were used without further purification.

### Synthesis of 19-nm Au nanospheres

The synthetic procedure of Au nanospheres was adapted from a previously published method.[10] Firstly, 5 mL of 0.2 M CTAB and 5 mL of 0.5 mM $HAuCl_4$ aqueous solutions were mixed at 25 °C under stirring, followed by the quick injection of 0.6 mL of 0.01 M ice-cold $NaBH_4$ aqueous solution. The mixture was under violent stirring for 2 min and then kept still for another 30 min. Secondly, 0.5 mL of the afore-mentioned mixture was mixed with 50 mg of CTAB, 0.05 mL of ethanolamine, 2 mL of 2.3 mM $HAuCl_4$ and 5 mL of DI water. The mixture was kept overnight until it turned to fresh red color.

### Synthesis of 14-nm Pd nanocubes

Pd nanocubes were synthesized according to a previously reported method.[38] Briefly, 105 mg of PVP, 60 mg of AA, 400 mg of KBr and 185 mg of KCl were dissolved in 8 mL of DI water. The solution was transferred to a vial, heated to 80 °C and kept for 10 min, followed by the addition of 3 mL of aqueous solution containing 57 mg of $Na_2PdCl_4$ with a pipette. The solution was kept at 80 °C for 10 min with a cap. After that, the final product was obtained by centrifugation at 13,000 rpm and washed with a mixture of DI water and ethanol for three times for further use.

### Synthesis of 7-nm CuPt bimetallic nanocubes

CuPt bimetallic nanocubes were generated through a previously reported method.[39] Briefly, 0.03 mmol of $K_2PtCl_4$, 0.03 mmol of $CuCl_2$, 9 mmol of KBr and 100 mg of PVP were dissolved in 15 mL of DI water, followed by the addition of 0.15 mL of 1 M HCl solution. The mixture was then transferred to a 25 mL autoclave and heated at 160 °C for 4 h. After cooling down to room temperature, the solution was centrifuged at 13,000 rpm and the final product was washed with DI water for three times for further use.

### Synthesis of Au and M@Au (M = Pd, CuPt) hybrid nanostars

Au nanostars were synthesized by a seed-mediated approach in an aqueous solution containing as-preformed Au seeds, CTAB, and $HAuCl_4$, with ethanolamine as both a capping and reducing agent. In a standard procedure, 0.1 mL of 19-nm Au nanosphere solution was mixed with 50 mg of CTAB, 0.05 mL of ethanolamine and 1 mL of 2.3 mM $HAuCl_4$ solution. Additional 5.95 mL of DI water was added to keep the overall volume to 7.1 mL. The mixture was stirred at 25 °C overnight to ensure the completion of the reaction. The final product was centrifuged at 5,000 rpm and washed with DI water for further characterization. For the synthesis of hybrid nanostars, the procedure was similar to that of Au nanostars except that Pd nanocubes or CuPt bimetallic nanocubes were used instead of Au nanospheres.

### Characterizations

Transmission electron microscopy (TEM) images were obtained with a Hitachi HT7700 at an accelerating voltage of 100 kV. Ultraviolet-visible (UV-*vis*) spectra were collected with a Shimadzu UV-3150 spectrometer under absorption mode.

### Surface-enhanced Raman spectroscopy

Au nanoparticles were re-dispersed in DI water and dropped on identical silicon wafers to form films. After the wafers were dried, 5 μL of $10^{-6}$ M rhodamine 6G ethanol solution was dropped and dispersed on the surface of the Au films. Raman signals were collected by a Bruker SENTERRA dispersive Raman microscope with excitation laser of 532 nm and 10 mW.

## Results and discussion

**Synthesis of Au nanostars**

**Figure 1a** shows a schematic illustration of the strategy used in synthesizing Au and M@Au (M = Pd, CuPt) hybrid nanostars. The nanostars were typically synthesized by reducing HAuCl$_4$ with ethanolamine in an aqueous solution containing as-preformed metallic NPs and CTAB serving as the seeds and stabilizer, respectively. Obviously, the experimental parameters such as precursor concentration and reaction temperature have a great influence on the shape of the Au NPs due to the different reaction kinetics. Using Au nanostars as an example, we characterized their shapes, structures, and LSPR properties, together with clarifying the growth mechanism presented in the synthesis. Au nanospheres of 19 nm in size that obtained by a two-step reduction procedure (see **Figure S1**) were used as seeds for the synthesis of Au nanostars. **Figure 1, b** and **c** show typical TEM images of the Au nanostars that obtained using the standard procedure (see Experimental for details). As observed from the TEM images, most of the Au NPs consisted of various arms (like stars). The as-prepared Au nanostars were uniform with a typical overall size of about 90 nm. To simplify the discussion in the next part of growth mechanism, we introduced the core size $D$ that defined as the maximum incircle diameter of a nanostar (**Figure 1c**). The Au NPs were used as the seeds in order to generate uniform nanostars. In the absence of Au seeds, Au nanostars with various sizes were generated while other parameters were kept the same (**Figure S2**).

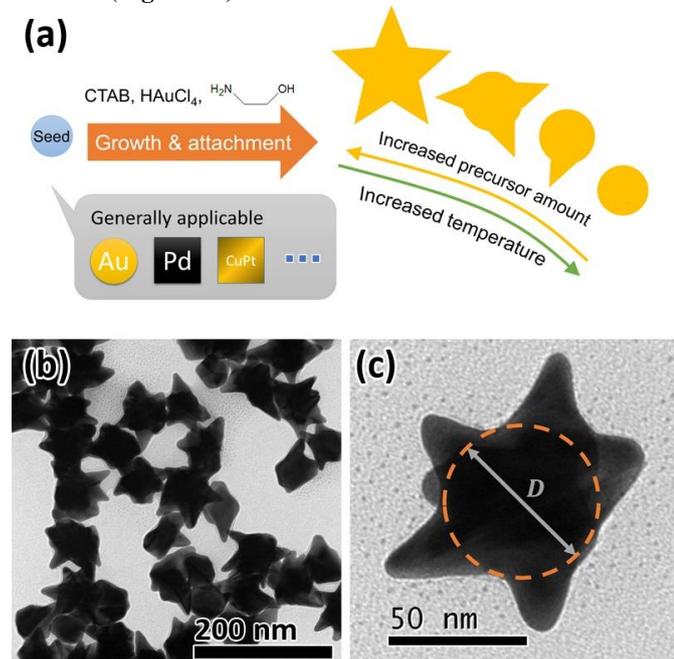

**Figure 1.** (a) Schematic illustration of controlled synthesis of Au and M@Au (M = Pd, CuPt) hybrid nanostars. (b, c) TEM images of Au nanostars obtained by the standard procedure. The dashed circle in (c) indicates the definition of the core size D.

Compared our synthetic procedure with those in the previous reports,[40-42] it is clear that the use of ethanolamine as a reducing agent is indispensable for the synthesis of the Au nanostars. The role of ethanolamine has been identified by replacing it with AA, which is a widely-used reducer for Au NPs. More specifically, 3 mg of AA instead of ethanolamine was used in the synthesis of Au seeds as well as in the overgrowth procedure of Au NPs while keeping other parameters unchanged. From the TEM images in **Figure S3**, only inhomogeneous Au NPs with various shapes were formed, indicating that ethanolamine plays a key role in facilitating a highly anisotropic growth of Au. It is well-known that ethanolamine is a weak base that contains hydroxyl and amino functional groups. The amino group has an appropriate power to reduce Au salt precursor. In addition, ethanolamine can also serve as a capping agent to selectively bind onto some specific planes of Au NPs by the amino group, and thereby promote the highly anisotropic growth.[32] More importantly, the alkaline condition can extensively suppress the intraparticle ripening induced by the chlorine ions that existed in the reaction system, and thus stabilize such branched Au nanostructures.[30] All of these characteristics are beneficial to the formation of Au nanostars. To better understand its role in the formation of Au nanostars, we further chose different types of amines in place of ethanolamine. Figure S4 a-c, shows TEM images of Au NPs that obtained in the presence of ethylenediamine, diethanolamine, and triethanolamine, respectively. Figure S4d shows the corresponding UV-*vis* spectra. From these TEM images, no Au nanostars were successfully observed. This demonstration was also supported by UV-*vis* spectra, in which only single LSPR peak was detected. It is clear that ethylenediamine is a stronger complex former than ethanolamine due to the two amino groups, thus much lower reaction rate and smaller Au NPs were obtained. Diethanolamine is a secondary amines while the hydroxyl group is retained, which results in weaker capping ability to the surface of Au due to its higher steric hindrance than ethanolamine. In this case, Au NPs with minor protuberances were achieved. Triethanolamine as a tertiary amine has even weaker capping ability, thus big inhomogeneous NPs were generated.

**Formation mechanisms and LSPR properties**

In order to further understand the process how Au nanostars were formed, we performed quantitative experiments with the aid of TEM observation and UV-*vis* monitoring. The formation mechanism was firstly studied by varying the amount of HAuCl$_4$ fed in the synthesis. For simplicity, samples that obtained by varying the amount of HAuCl$_4$ solution from 0.2 to 0.4, 0.6, 0.8, and 1.0 mL were denoted as Au-p20, Au-p40, Au-p60, Au-p80, and Au-p100, respectively. TEM images and UV-*vis* spectra of these samples are shown in **Figure 2** and **Figure 3a**, respectively. From the TEM images in **Figure 2**, we can observe that the overall size of the Au NPs increased along with the amount of HAuCl$_4$. In addition, the morphology of the Au NPs gradually evolved into branched nanostructures with more anisotropic characters. **Figure 3**, **b** and **c** show strongly positive correlation of core size (as defined in **Figure 1c**) and average observed number of branches per nanostar with the amount of gold precursor, respectively. The core size of the nanostars was gradually increased from 19 nm (as-preformed Au seeds) to 53 nm when 1.0 mL of $HAuCl_4$ solution was added. Meanwhile, the observed number of Au branches increased to over 2.5 per nanostar. As such, these TEM observations indicate that the nanostars were generated through the overgrowth of the core in combination with the anisotropic growth of the branches.

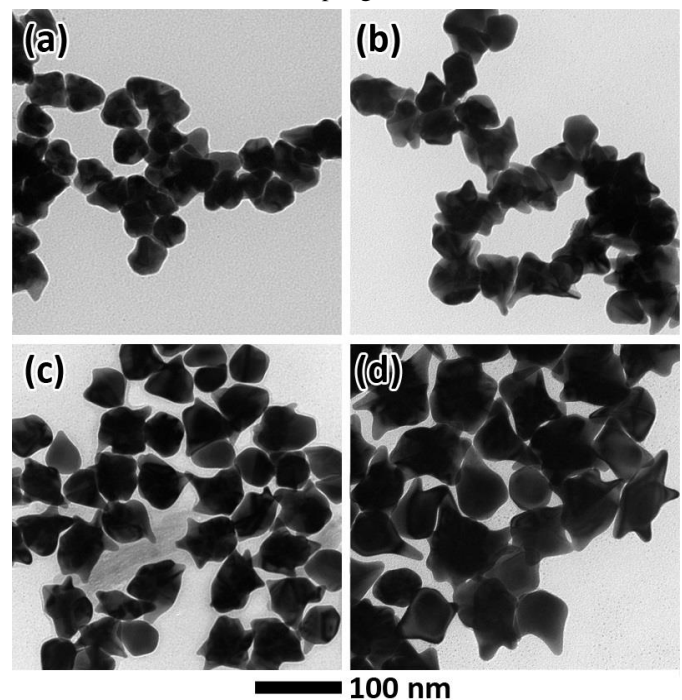

**Figure 2.** TEM images of (a) Au-p20, (b) Au-p40, (c) Au-p60 and (d) Au-p80 samples. Images share a scale bar of 100 nm.

Such morphology evolution of the Au NPs dependent on the amount of gold precursor was also monitored by their LSPR properties from UV-*vis* absorption spectra (**Figure 3a**). Since the LSPR of Au NPs can be highly affected by surfactant and the surrounding dielectric environment (*e.g.*, solvent, capping agent, etc.),[18] the concentration of CTAB was kept the same in all experiments. From **Figure 3a**, the Au nanostars that obtained by the standard procedure (1.0 mL of $HAuCl_4$ was used) show a primary absorption peak at 734 nm, together with a shoulder in the range of 500 to 600 nm. Obviously, this shoulder refers to the LSPR from the core of the nanostars, which represents the similar resonance mode as that in spherical Au seeds at 523 nm.[40] While, the primary absorption peak in near-IR could be attributed to the longitudinal resonance mode from the branches. The transverse mode from the branches was not distinguished possibly because they are too weak compared to that shown by the core, which also fell in the visible range. This demonstration was confirmed by the UV-*vis* spectra collected from the samples with different amount of $HAuCl_4$. The near-IR absorption peak raised gradually when the amount of $HAuCl_4$ increased, indicating the formation of nanostructures with higher aspect ratio, *i.e.* the overgrowth of Au with anisotropic morphology. The intensity ratio of the two peaks dependent on precursor amount is also shown in **Figure 3c**, which is in good agreement with the correlation shown by observed branch number from TEM observation. Meanwhile, the peak wavelength in the range of 500 to 600 nm also red-shifted when the core size increased with the amount of gold precursor.

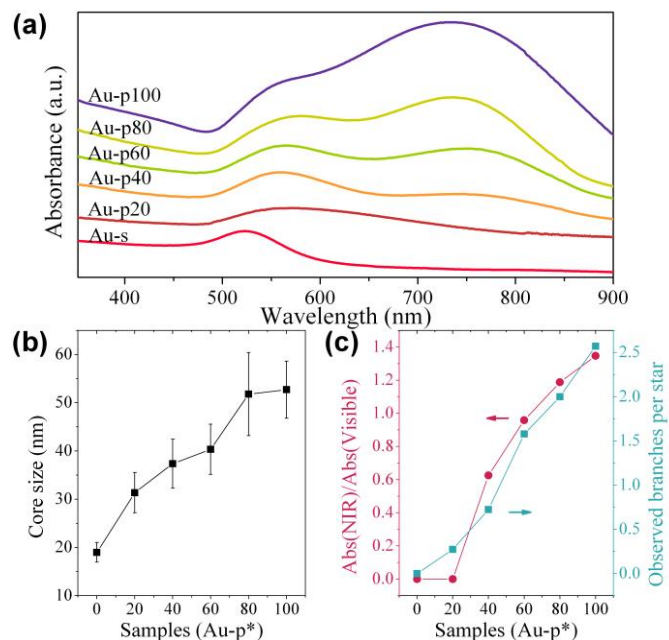

**Figure 3.** (a) UV-*vis* spectra of Au nanostars prepared with different amount of $HAuCl_4$, including the as-performed Au spherical seed. Spectra are offset for clarity. (b) Average core size of these different Au nanostar samples. Error bars indicate standard deviation. (c) Peak intensity ratio of absorbance in near-IR (NIR) and visible ranges (red discs, left axis) and observed number of branches per nanostar from TEM images (cyan squares, right axis) in these different Au nanostar samples. In (b) and (c), sample Au-s appears as Au-p0.

We attribute these phenomena to the concurrence and competition of two formation mechanisms. First, growing of Au adatoms on Au seeds follows a conventional epitaxial mechanism (*i.e.*, Frank-van der Merwe growth mode)[43] since the contact angle of Au on Au seeds is zero (mechanism I). The shift of visible light peaks reflects such growth of the core, which was also revealed by the increase of the core size when precursor amount was increased. This demonstration was also supported by TEM imaging the products obtained at different reaction times in the initial stage (**Figure S5**). As observed, the Au cores gradually grew up with the reaction time, showing a linear relation. In the seeded growth of heterogeneous nanocrystals, there are generally three types of growth modes that are responsible for the final products. Obviously, epitaxial growth mechanism results in the core-shell nanostructures. In this homogeneous growth, the epitaxial growth mechanism only leads to the increase of the Au core in size. The similar results were also observed in the heterogeneous growth of Pd@Au and CuPt@Au nanostructures, and would be discussed in the next section. Second, rising of the near-IR peak represents the formation of Au branches, which possibly results from the attachment of small nanoclusters onto the nanoparticles (mechanism II). Careful observation shows that many nanometer-sized clusters co-existed with the nanostars, for example in **Figure 1c,** which confirms our demonstration. When precursor concentration was increased, higher proportion of gold atoms are nucleated into nanoclusters compared to that grown on seeds, which facilitated the formation of branches. This demonstration was further supported by the morphology evolution of the aged samples after the reaction was completed. After being aged at 0 °C, the branches on Au nanostars became much longer as indicated in **Figure S6**, with nanoclusters disappeared. This result indicates that the attachment mechanism contributes to the formation of the branches.

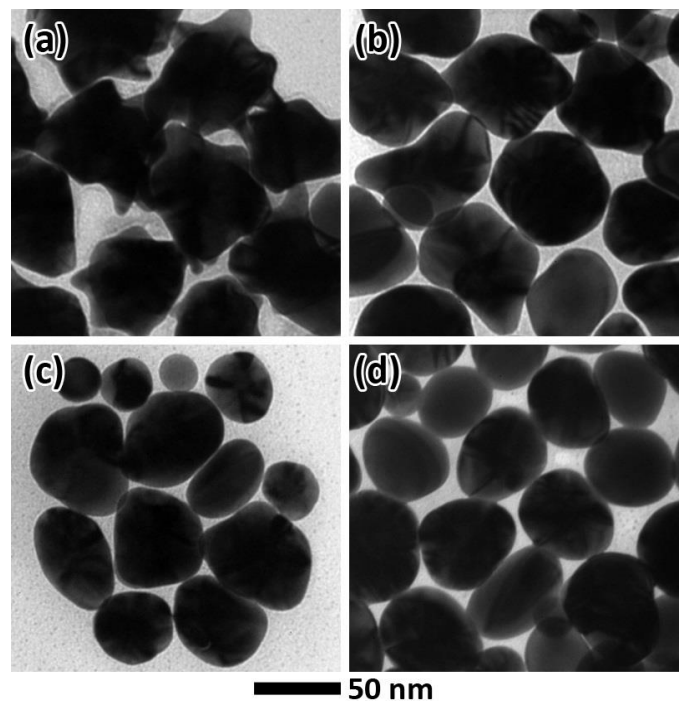

**Figure 4.** TEM images of Au samples prepared using the standard procedure except for different reaction temperature (a) 35 °C, (b) 50 °C, (c) 65 °C and (d) 80 °C. Images share a scale bar of 50 nm.

We have also systematically investigated the effect of reaction temperature on the morphology of the Au NPs and the corresponding LSPR properties. **Figure 4** shows TEM images of the Au NPs that synthesized at different reaction temperature while keeping other conditions unchanged. As observed, low temperature (more specifically below 50 °C) facilitated the formation of branches. With the reaction temperature increased, however, the as-obtained samples show spherical morphology instead. This result was also confirmed by the corresponding UV-*vis* spectra as shown in **Figure 5**. Samples synthesized at different temperature ranging from 35 to 95 °C are denoted as Au-t35 to Au-t95. It is clear that relatively higher temperature (higher than 50 °C) leads to the disappearance of the near-IR peak associated with the anisotropic structure. The inset of **Figure 5** shows the relationship between the visible peak wavelength and reaction temperature, where these peaks blue-shifted with temperature increased. This blue shift might result from the decreased particle size (**Figure S7**) as well as the aspect ratio in the specified resonance mode, which also implies that the particles became more spherical. It is well-known that elevated temperature can accelerate the diffusion of atoms and then maintain the adatoms around the seeds at a sufficiently high level, thereby leading to the conformal growth.[44] In addition, previous studies showed that the intraparticle diffusion and ripening associated with reaction temperature might also be responsible for the disappearance of anisotropic morphology.[30, 45]

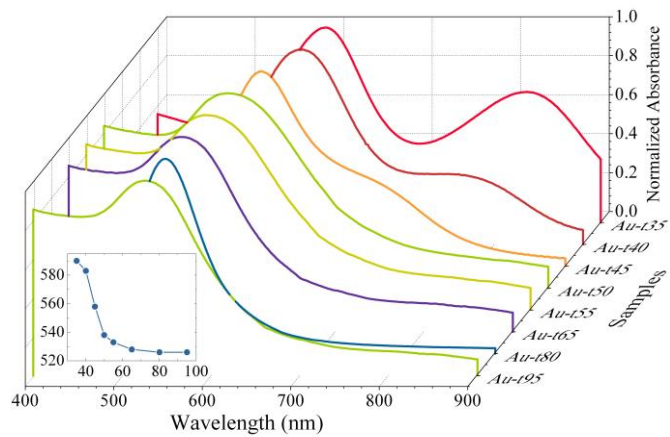

**Figure 5.** UV-*vis* spectra for temperature series samples. Inset shows the relationship of core peak wavelength (nm) versus synthesis temperature (°C).

**Core@shell hybrid structure**

Our synthetic approach can be also extended to the synthesis of core@shell hybrid nanostructures using 14-nm Pd and 7-nm CuPt nanocubes as the seeds. These heterogeneous seeds were synthesized by previously reported methods with modification described in the Experimental. The corresponding TEM images are shown in **Figure S8.** Similar to the homogeneous growth with Au nanospheres as the seeds, the amount of $HAuCl_4$ has a great influence on the morphology of hybrid nanostructures *via* heterogeneous growth (see **Figures 6 and 7**). With Pd nanocubes as the seeds, for example, Pd@Au hybrid nanospheres were obtained when the ratio of $HAuCl_4$ and the seeds was equivalent to 1 (**Figure 6a**). The Moire fringes can be observed in the interior of Pd@Au hybrid nanospheres due to the lattice mismatch between Au and Pd, indicating the epitaxial growth of Au shell on Pd seeds. This result further confirms the involvement of the F-M mechanism in the formation of core-shell nanostrucutrues.[46,47] When the ratio of $HAuCl_4$ and the seeds was increased to 12, branched Pd@Au hybrid nanostructures were generated through the conventional layered growth followed by the random attachment of Au nanoclusters (**Figure 6b**). The LSPR properties of the Pd@Au hybrid nanostructures were also tailored by the shape of Au shells as shown in **Figure 6c**. In addition to the slightly red-shifted visible peak, the branched nanostructures also show an additional LSPR peak at near-IR region in comparison with the spherical hybrid nanostructures due to the anisotropic characteristics. This result is in agreement with the case of Au. When using CuPt nanocubes as the seeds, ternary CuPt@Au hybrid nanostructures with both spherical and branched shapes were also generated by varying the ratio of $HAuCl_4$ and the seeds as shown in **Figure 7**, **a** and **b**. These hybrid nanostructures exhibit similar LSPR properties as that of Au and Pd@Au (**Figure 7c**). This work not only provides a facile and versatile strategy to synthesize Au and Au-based hybrid nanostars, but also makes it possible to fabricate multifunctional nanostructures with distinct core metals, *e.g.*, in the application of light-enhanced catalysis enabled by the introduction of a plasmonic material to noble metal catalysts.[48,49]

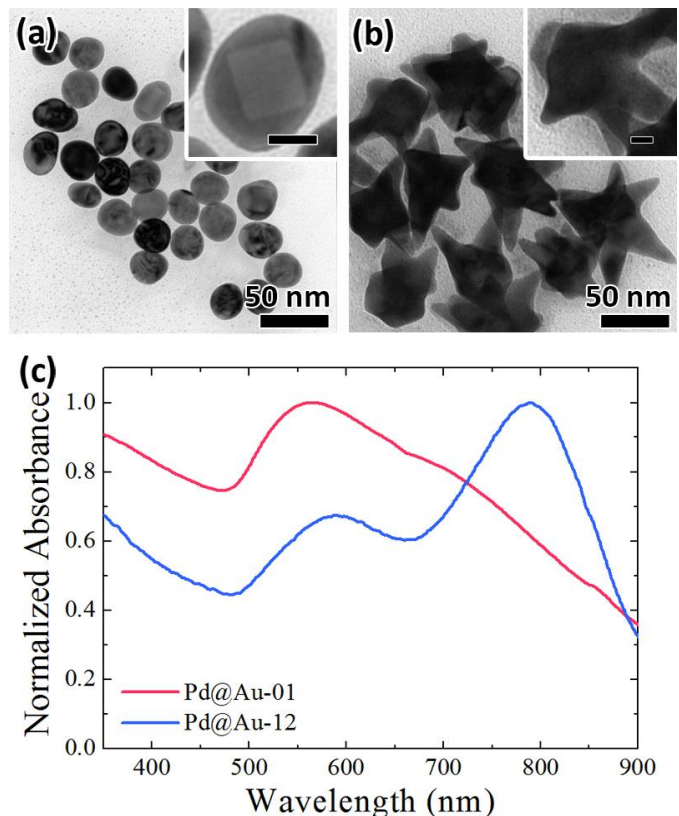

**Figure 6.** (a, b) TEM images of (a) Pd@Au-01 and (b) Pd@Au-12 samples. Insets show single particles and scale bar is 10 nm. (c) UV-*vis* spectra for the two Pd@Au samples. The absorbance has been normalized for comparing.

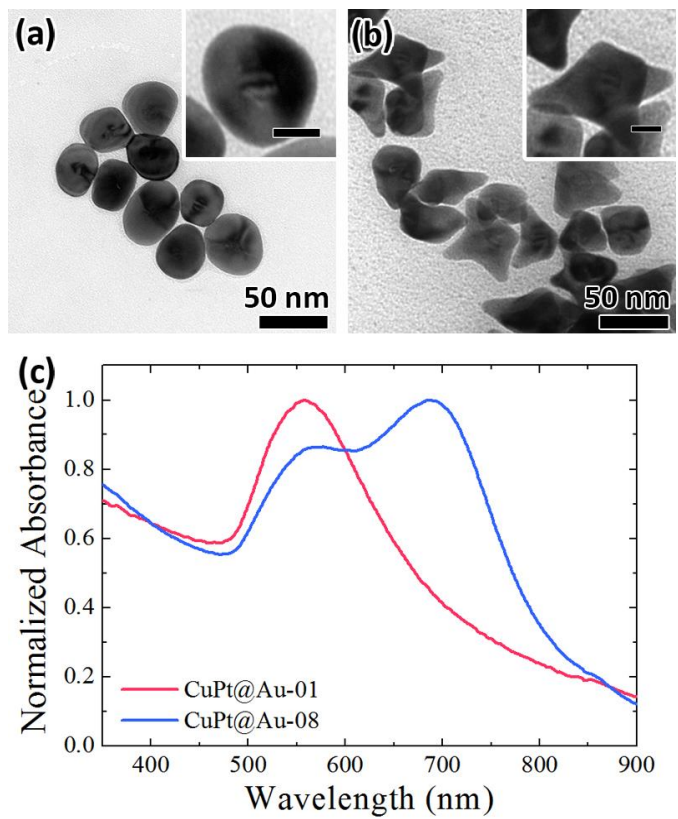

**Figure 7.** (a, b) TEM images of (a) CuPt@Au-01 and (b) CuPt@Au-08 samples. Insets show single particles, scale bar 10 nm. (c) UV-*vis* spectra for the two CuPt@Au samples. The absorbance has been normalized for comparing.

**SERS performance**

The as-prepared Au-based nanostructures were evaluated as the substrate for SERS. We firstly deposited Au films on silicon wafers by drop-casting aqueous dispersion of the Au NPs solution. After the films were dried in ambient environment, 5 μL of $10^{-6}$ M rhodamine 6G ethanol solution was dropped onto the wafers and dried. Raman spectra collected on different samples including Au nanostar films, Au nanosphere film and bare silicon wafer are shown in **Figure 8**. On the bare silicon wafer without Au film, almost no Raman signal of rhodamine 6G was detected. The addition of Au films dramatically enhanced the Raman signals of rhodamine 6G due to the larger localized electric field. Owing to the sharp edges and tips associated with branched structures,[18, 19] and the Au nanostars (*e.g.*, the samples Au-p40 and Au-p100) gave two-fold enhancement of SERS signals than the Au nanospheres.

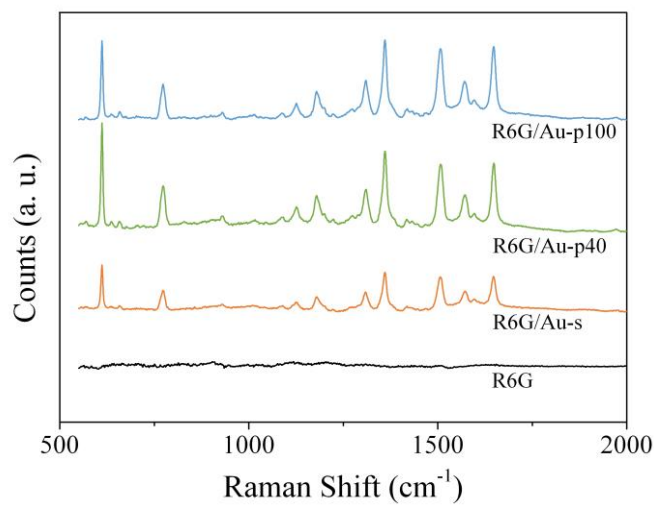

**Figure 8.** Surface-enhanced Raman spectra of rhodamine 6G obtained on different Au substrates on Si wafers. Traces are offset for clarity.

## Conclusions

We have developed a general and facile strategy to synthesize Au nanostars as well as M@Au (M = Pd and CuPt for example) hybrid nanostars in an aqueous solution. Ethanolamine was found to play a key role in facilitating the formation of the branched nanostructures serving as both a reducing and capping agent. In addition, the precursor amount and reaction temperature also have a great influence on the morphology of the Au nanostructures. TEM observation and UV-*vis* monitoring indicate that two concurrent mechanisms are both responsible for the formation of the Au nanostars. During this synthesis, the cores grow up by the layered epitaxial mechanism (*i.e.*, Frank-van der Merwe growth mode), while, the random attachment of Au nanoclusters onto the cores accounts for the formation of the branches. The LSPR properties of the Au nanostructures are highly sensitive to their morphology, which were simply controlled by the precursor amount and reaction temperature. The as-prepared Au nanostars show improved SERS performance in detecting trace amount of rhodamine 6G compared to that of the Au nanospheres due to their sharp edges and tips.

## Acknowledgements

This work was supported by NSFC (No. 51372222) and the Fundamental Research Funds for the Central Universities (No. 2014FZA4007). J Du and J Yu also acknowledge the financial support from National University Student Innovation Project (No. 201310335067) from Ministry of Education of China and Zhejiang University.

## Notes and references

State Key Laboratory of Silicon Materials, Department of Materials Science and Engineering and Cyrus Tang Center for Sensor Materials and Applications, Zhejiang University, Hangzhou 310027, China
Email: msezhanghui@zju.edu.cn; Fax: +86 571 87952322; Tel: +86 571 87953190

1. R. M. Crooks, M. Zhao, L. Sun, V. Chechik and L. K. Yeung, *Acc. Chem. Res.,* 2000, **34**, 181-190.
2. K. L. Kelly, E. Coronado, L. L. Zhao and G. C. Schatz, *J. Phys. Chem. B*, 2002, **107**, 668-677.
3. E. Boisselier and D. Astruc, *Chem. Soc. Rev.,* 2009, **38**, 1759-1782.
4. N. Li, P. Zhao and D. Astruc, *Angew. Chem. Int. Ed.*, 2014, **53**, 1756-1789.
5. H. Chen, L. Shao, Q. Li and J. Wang, *Chem. Soc. Rev.,* 2013, **42**, 2679-2724.
6. P. K. Jain, X. H. Huang, I. H. El-Sayed and M. A. El-Sayed, *Acc. Chem. Res.,* 2008, **41**, 1578-1586.
7. C. J. Murphy, A. M. Gole, J. W. Stone, P. N. Sisco, A. M. Alkilany, E. C. Goldsmith and S. C. Baxter, *Acc. Chem. Res.,* 2008, **41**, 1721-1730.
8. L. Vigderman, B. P. Khanal and E. R. Zubarev, *Adv. Mater.,* 2012, **24**, 4811-4841, 5014.
9. J. X. Gao, C. M. Bender and C. J. Murphy, *Langmuir*, 2003, **19**, 9065-9070.
10. B. Nikoobakht and M. A. El-Sayed, *Chem. Mater.,* 2003, **15**, 1957-1962.
11. A. Gole and C. J. Murphy, *Chem. Mater.,* 2004, **16**, 3633-3640.
12. X. H. Huang, S. Neretina and M. A. El-Sayed, *Adv. Mater.,* 2009, **21**, 4880-4910.
13. C. L. Nehl, H. Liao and J. H. Hafner, *Nano Lett.,* 2006, **6**, 683-688.
14. P. S. Kumar, I. Pastoriza-Santos, B. Rodríguez-González, F. J. García de Abajo and L. M. Liz-Marzán, *Nanotechnology*, 2008, **19**, 015606.
15. A. Guerrero-Martínez, S. Barbosa, I. Pastoriza-Santos and L. M. Liz-Marzán, *Curr. Opin. Colloid Interface Sci.*, 2011, **16**, 118-127.
16. A. A. Umar and M. Oyama, *Cryst. Growth Des.*, 2009, **9**, 1146-1152.
17. N. Ortiz and S. E. Skrabalak, *Cryst. Growth Des.*, 2011, **11**, 3545-3550.
18. S. Eustis and M. A. El-Sayed, *Chem. Soc. Rev.,* 2006, **35**, 209-217.
19. E. Hao, R. C. Bailey, G. C. Schatz, J. T. Hupp and S. Li, *Nano Lett.,* 2004, **4**, 327-330.
20. K. A. Willets and R. P. Van Duyne, *Annu. Rev. Phys. Chem.,* 2007, **58**, 267-297.
21. S. Barbosa, A. Agrawal, L. Rodríguez-Lorenzo, I. Pastoriza-Santos, R. A. Alvarez-Puebla, A. Kornowski, H. Weller and L. M. Liz-Marzán, *Langmuir*, 2010, **26**, 14943-14950.
22. A. Kedia and P. S. Kumar, *J. Mater. Chem. C*, 2013, **1**, 4540-4549.
23. A. Kedia and P. S. Kumar, *J. Phys. Chem. C*, 2012, **116**, 1679-1686.
24. T. K. Sau, A. L. Rogach, M. Doblinger and J. Feldmann, *Small*, 2011, **7**, 2188-2194.
25. L.-C. Cheng, J.-H. Huang, H. M. Chen, T.-C. Lai, K.-Y. Yang, R.-S. Liu, M. Hsiao, C.-H. Chen, L.-J. Her and D. P. Tsai, *J. Mater. Chem.*, 2012, **22**, 2244-2253.
26. B. L. Sanchez-Gaytan, P. Swanglap, T. J. Lamkin, R. J. Hickey, Z. Fakhraai, S. Link and S.-J. Park, *J. Phys. Chem. C*, 2012, **116**, 10318-10324.
27. S. Umadevi, H. C. Lee, V. Ganesh, X. Feng and T. Hegmann, *Liq. Cryst.,* 2013, **41**, 265-276.
28. P. Pallavicini, G. Chirico, M. Collini, G. Dacarro, A. Dona, L. D'Alfonso, A. Falqui, Y. Diaz-Fernandez, S. Freddi, B. Garofalo, A.

Genovese, L. Sironi and A. Taglietti, *Chem. Commun.,* 2011, **47**, 1315-1317.
29. A. Casu, E. Cabrini, A. Dona, A. Falqui, Y. Diaz-Fernandez, C. Milanese, A. Taglietti and P. Pallavicini, *Chem. Eur. J.*, 2012, **18**, 9381-9390.
30. L. Zhao, X. Ji, X. Sun, J. Li, W. Yang and X. Peng, *J. Phys. Chem. C*, 2009, **113**, 16645-16651.
31. G. Maiorano, L. Rizzello, M. A. Malvindi, S. S. Shankar, L. Martiradonna, A. Falqui, R. Cingolani and P. P. Pompa, *Nanoscale*, 2011, **3**, 2227-2232.
32. A.-J. Wang, Y.-F. Li, M. Wen, G. Yang, J.-J. Feng, J. Yang and H.-Y. Wang, *New J. Chem.,* 2012, **36**, 2286.
33. W. Jia, J. Li and L. Jiang, *ACS Appl. Mat. Interfaces* 2013, **5**, 6886-6892.
34. C. J. Serpell, J. Cookson, D. Ozkaya and P. D. Beer, *Nat. Chem.*, 2011, **3**, 478-483.
35. S. Zhou, B. Varughese, B. Eichhorn, G. Jackson and K. McIlwrath, *Angew. Chem.*, 2005, **117**, 4615-4619.
36. J. K. Nørskov, F. Abild-Pedersen, F. Studt and T. Bligaard, *Proc. Nat. Acad. Sci.*, 2011, **108**, 937-943.
37. C. J. DeSantis, R. G. Weiner, A. Radmilovic, M. M. Bower and S. E. Skrabalak, *J. Phys. Chem. Lett.*, 2013, **4**, 3072-3082.
38. M. Jin, H. Liu, H. Zhang, Z. Xie, J. Liu and Y. Xia, *Nano Res.*, 2011, **4**, 83-91.
39. A. X. Yin, X. Q. Min, W. Zhu, W. C. Liu, Y. W. Zhang and C. H. Yan, *Chem. Eur. J.*, 2012, **18**, 777-782.
40. N. R. Jana, L. Gearheart and C. J. Murphy, *Langmuir*, 2001, **17**, 6782-6786.
41. C. Ziegler and A. Eychmüller, *J. Phys. Chem. C*, 2011, **115**, 4502-4506.
42. C. Gao, J. Vuong, Q. Zhang, Y. Liu and Y. Yin, *Nanoscale*, 2012, **4**, 2875-2878.
43. E. Bauer and J. H. van der Merwe, *Phys. Rev. B*, 1986, **33**, 3657-3671.
44. J. Zeng, C. Zhu, J. Tao, M. Jin, H. Zhang, Z. Y. Li, Y. Zhu and Y. Xia, *Angew. Chem. Int. Ed.*, 2012, **51**, 2354-2358.
45. L. Rodríguez-Lorenzo, J. M. Romo-Herrera, J. Pérez-Juste, R. A. Alvarez-Puebla and L. M. Liz-Marzán, *J. Mater. Chem.*, 2011, **21**, 11544.
46. F. R. Fan, D. Y. Liu, Y. F. Wu, S. Duan, Z. X. Xie, Z. Y. Jiang and Z. Q. Tian, *J. Am. Chem. Soc.*, 2008, **130**, 6949-6951.
47. M. Tsuji, D. Yamaguchi, M. Matsunaga and M. J. Alam, *Cryst. Growth Des.,* 2010, **10**, 5129-5135.
48. S. Sarina, H. Zhu, E. Jaatinen, Q. Xiao, H. Liu, J. Jia, C. Chen and J. Zhao, *J. Am. Chem. Soc.*, 2013, **135**, 5793-5801.
49. X. Huang, Y. Li, Y. Chen, H. Zhou, X. Duan and Y. Huang, *Angew. Chem. Int. Ed.*, 2013, **52**, 6063-6067.